\documentclass[sn-mathphys,Numbered]{sn-jnl}

\usepackage{graphicx}%
\usepackage{multirow}%
\usepackage{amsmath,amssymb,amsfonts}%
\usepackage{amsthm}%
\usepackage{mathrsfs}%
\usepackage[title]{appendix}%
\usepackage{xcolor}%
\usepackage{textcomp}%
\usepackage{manyfoot}%
\usepackage{booktabs}%
\usepackage{algorithm}%
\usepackage{algorithmicx}%
\usepackage{algpseudocode}%
\usepackage{listings}%

\theoremstyle{thmstyleone}%
\theoremstyle{thmstyletwo}%

\theoremstyle{thmstylethree}%

\begin{document}

\title[News ecosystem dynamics: Supply, Demand, Diffusion, and the role of Disinformation]{News ecosystem dynamics: Supply, Demand, Diffusion, and the role of Disinformation}


\author*[1,3]{\fnm{Pietro} \sur{Gravino}}\email{pietro.gravino@sony.com}

\author[1,3]{\fnm{Giulio} \sur{Prevedello}}\email{giulio.prevedello@sony.com}

\author[2,3]{\fnm{Emanuele} \sur{Brugnoli}}\email{emanuele.brugnoli@sony.com}

\affil[1]{\orgdiv{Sony CSL}, \orgname{Paris Research}, \orgaddress{\street{6, Rue Amyot}, \city{Paris}, \postcode{75005}, \country{France}}}

\affil[2]{\orgdiv{Sony CSL}, \orgname{Rome Research, Joint Initiative CREF-SONY, Centro Ricerche Enrico Fermi}, \orgaddress{\street{Via Panisperna 89/A}, \city{Rome}, \postcode{00184}, \country{Italy}}}

\affil[3]{\orgname{Enrico Fermi’s Research Center}, \orgaddress{\street{Via Panisperna 89/A}, \city{Rome}, \postcode{00184}, \country{Italy}}}

\abstract{The digital age provides new challenges as information travels more quickly in a system of increasing complexity. 
But it also offers new opportunities, as we can track and study the system more efficiently.
Several studies individually addressed different digital tracks, focusing on specific aspects like disinformation production or content-sharing dynamics.
In this work, we propose to study the news ecosystem as an information market by analysing three main metrics: Supply, Demand, and Diffusion of information.
Working on a dataset relative to Italy from December 2019 to August 2020, we validate the choice of the metrics, proving their static and dynamic relations, and their potential in describing the whole system.
We demonstrate that these metrics have specific equilibrium relative levels.
We reveal the strategic role of Demand in leading a non-trivial network of causal relations.
We show how disinformation news Supply and Diffusion seem to cluster among different social media platforms.
Disinformation also appears to be closer to information Demand than the general news Supply and Diffusion, implying a potential danger to the health of the public debate.
Finally, we prove that the share of disinformation in the Supply and Diffusion of news has a significant linear relation with the gap between Demand and Supply/Diffusion of news from all sources.
This finding allows for a real-time assessment of disinformation share in the system. 
It also gives a glimpse of the potential future developments in the modelisation of the news ecosystem as an information market studied through its main drivers.
}




\maketitle

\section{Introduction}\label{intro}

Internet and social media have significantly transformed how people access, share, and consume information.
While digital environments have considerably promoted disintermediation, enabling diverse voices to participate in the collective dialogue at the expense of professional information, the role of leader nodes in social networks (i.e., the main influential accounts) remains crucial in determining how information is disseminated and consumed~\cite{acampa2022,rehman2020,welbers2018}.
Recent works on the dynamics of information dissemination and consumption have surged interest in the complexity of information ecosystems, particularly focusing on disinformation from its very definition~\cite{kapantai2021,lazer2018} to its spread~\cite{delvicario2016} and connection to partisanship~\cite{garrett2021,pennycock2019}.
A significant portion of research has examined the impact of disinformation on human behaviour~\cite{bastick2021}, political elections~\cite{morgan2018}, sustainability~\cite{treen2020}, and health~\cite{sasahara2021}.
The term 'Infodemic'~\cite{simon2023}, which resurfaced during the Covid-19 pandemic~\cite{cinelli2020b}, describes the overwhelming flood of both accurate and false information about the virus, leading to confusion and harmful behaviours that exacerbated the pandemic~\cite{rocha2023}.
These investigations have led to questions about identifying statistical indicators in news content and consequently effective strategies for preventing the spread of disinformation~\cite{delvicario2019,guay2023,pacheco2020}. 

The broader information ecosystem, which includes both news producers (also referred as leaders) and consumers, has received less attention than disinformation itself.
Few attempts have been made to study the dynamics of interaction between news producers and consumers~\cite{elliot1998}, and we still lack a fundamental understanding of the system.
A previous work~\cite{gravino2022} identified the supply on the production side and the demand on the consumption side as the main drivers of the systemic dynamics.


The Supply, as the content produced by news producers, directly influences the demand side. Consumers are more likely to engage with and demand content that aligns with their interests, preferences, and values~\cite{cinelli2020a}. At the same time, the type and nature of news content also impact how widely it is shared and diffused. Engaging content significantly influences diffusion as followers are more likely to share relevant information~\cite{turcotte2015}.

The Demand is the needs and interests of consumers and influences the type and nature of news content that news producers create. Producers are likely to cater to topics and formats that align with the interests and demands of their audience~\cite{thurman2019}. At the same time, the demand for certain types of content influences how widely it is shared. If a particular piece of news resonates strongly with the audience, it is more likely to be widely diffused by followers~\cite{thompson2020}.

A third layer can be added to the analysis by looking at the Diffusion of contents, which can be defined as the sharing volume of news content. It influences future content creation strategies by news producers. Producers may observe what types of content are gaining traction through diffusion and adjust their production accordingly~\cite{andrews2010}. At the same time, the diffusion of news content affects what other consumers are exposed to and, consequently, what they may demand. Popular content that has been widely diffused may generate increased demand from new audiences~\cite{iyengar2004}. On the other hand, sharing behaviour can sometimes result in the formation of echo chambers, i.e. user groups that share a common narrative~\cite{brugnoli2019,pratelli_entropy-based_2023}, in persistent recurring patterns~\cite{desiderio_recurring_2023}, or in self-organised collective actions~\cite{mancini_self-induced_2022}. 

Separately, Supply, Demand and Diffusion have been subjects of several studies \cite{Brugnoli2023,patuelli_sustainable_2023,mattei_bow-tie_2022,clemm_von_hohenberg_truth_2023}, but together, they could provide a deeper and systemic understanding of the news ecosystem. 
Still, it has to be proved that their mutually influenced interplay underscores the complex dynamics of the news ecosystem.

Our study connects the dots between news Demand, Supply, and Diffusion, analysing the news ecosystem as a single complex system. 
We aim to prove that the chosen metrics effectively track the system and account for the phenomenology. 
Finally, we show how this approach helps better understand the system's health status, assessing disinformation production and spreading levels.
This work aims to confirm and generalise some of the results that emerged in a previous study~\cite{gravino2022}.
The previous observations will be expanded to include more keywords and different social networks, and model-agnostic techniques will be adopted to provide further generalisation.
We select the main news outlets in Italy, encompassing a wide range of news media outlets active from December 2019 to August 2020.
We monitor their posts' production and users' sharing volumes concerning the most relevant keywords in the observed period.
These keywords have been identified by looking at the most important keywords used in the Google Search Engine, which has also been used to track the Demand for information.
The dynamics of these interactions can differ among social media platforms because of variations in their business models and content selection algorithms~\cite{demarzo2023recommender,tommasel2022}.
For these reasons, we focused on the two main social media in the considered time frame: Facebook and X.
The latter will be referred to as Twitter, as this was still the name at the time of data gathering.


\section{Results and Discussion}\label{res}

In this work the information ecosystem is studied as a market driven by three main metrics: the Supply, the Demand, and the Diffusion of information. We will show how these quantities are related in terms of scales and dynamics without assuming any specific model. Then, we will show how the relation between these forces can be used to provide useful insights about the health status of the information system, providing an independent assessment of the Non-Trustworthy levels of information supplied and diffused.

\subsection{The three forces}\label{3forces}

\begin{figure}[!ht]
\centering
\includegraphics[width=1.\textwidth]{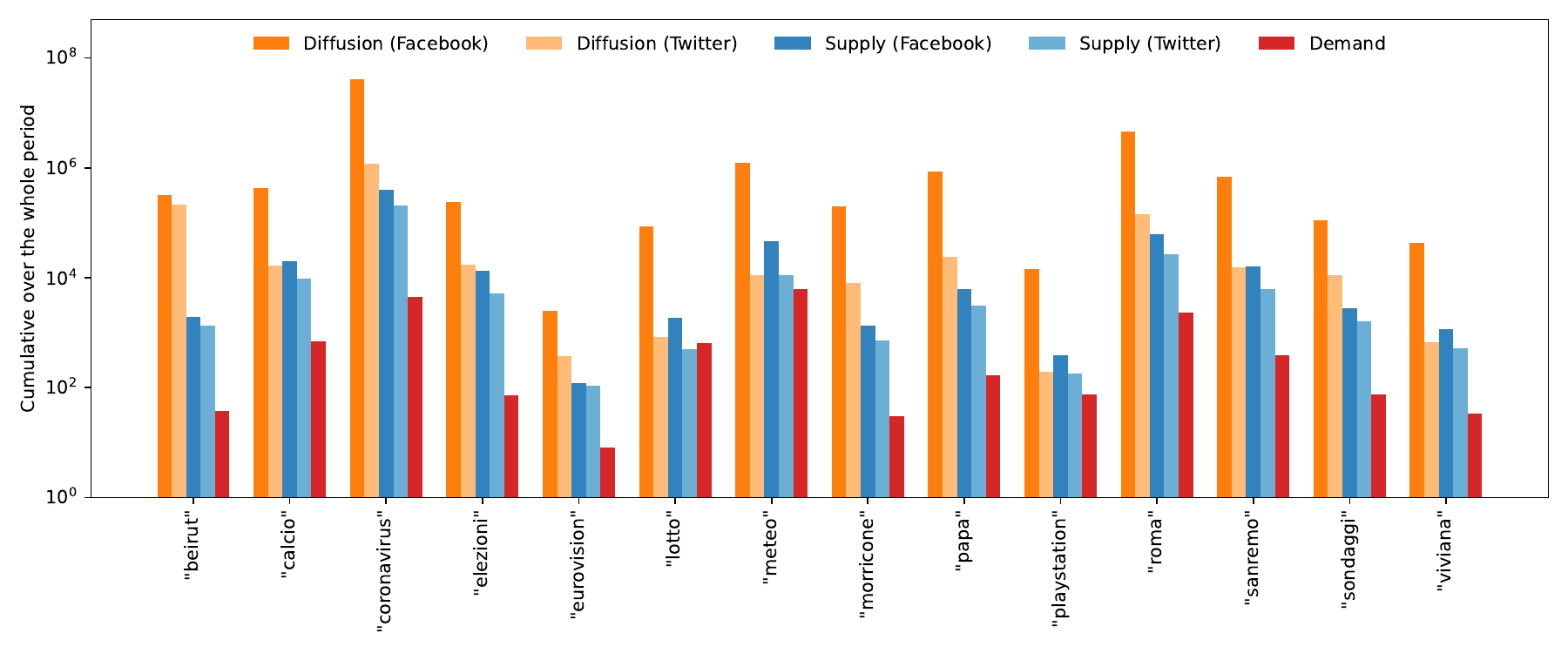}
\caption{The cumulative count over the whole period (from December '19 to August '20) of the three forces where Supply and Diffusion are reported for both Facebook (FB) and Twitter(TW) for all the keywords. The order of the forces is almost always the same for all keywords.}
\label{fig:overallhist}
\end{figure}

First, we formally introduce the three forces that are the main subjects of our investigation. 
Demand represents the aggregated need for information in the community. 
As a proxy for Demand, the daily time-series of keyword searches on Google has been collected from the Google Trends platform with a procedure that allowed us to elaborate an absolute scale valid for comparisons between different keywords (see Section~\ref{matmet}).
Supply represents the aggregated production of the most important news outlets.
We elaborate the information Supply by aggregating the posts' publication from an extensive list of news outlets' profiles on the two main social media (Facebook and Twitter), which is strictly linked with their overall news production (see Section~\ref{matmet}).
Diffusion represents the aggregated reaction of the community to the news posts.
We calculated the diffusion by summing the shares of the news items in the supply for both the two monitored social media (see Section~\ref{matmet}).
All forces are monitored for the most prominent keywords in Italy from December 2019 to August 2020. More details are provided in the Section \ref{matmet}.
We report the cumulative sum of the three forces over the monitored period in Fig.~\ref{fig:overallhist}.
Unless differently specified, the same force on different social media will be treated in the analysis as two different forces for two reasons. 
First, the two different social media are used by different communities in different ways, and we want to give an account about that.
Second, if the relations between the forces are the same and are independent of the platform, we should observe this, so treating them independently will serve as validation of the existence of deeper relations between the forces.
While the magnitude of the cumulative sum varies between the different keywords, the order of the cumulative sum of forces (from the larger to the smaller) for a given keyword is almost always the same.
This suggests the existence of relations between the forces.

\subsection{Correlations and relative scale}\label{corr eq}
\begin{figure}[!ht]
\centering
\includegraphics[width=.75\textwidth]{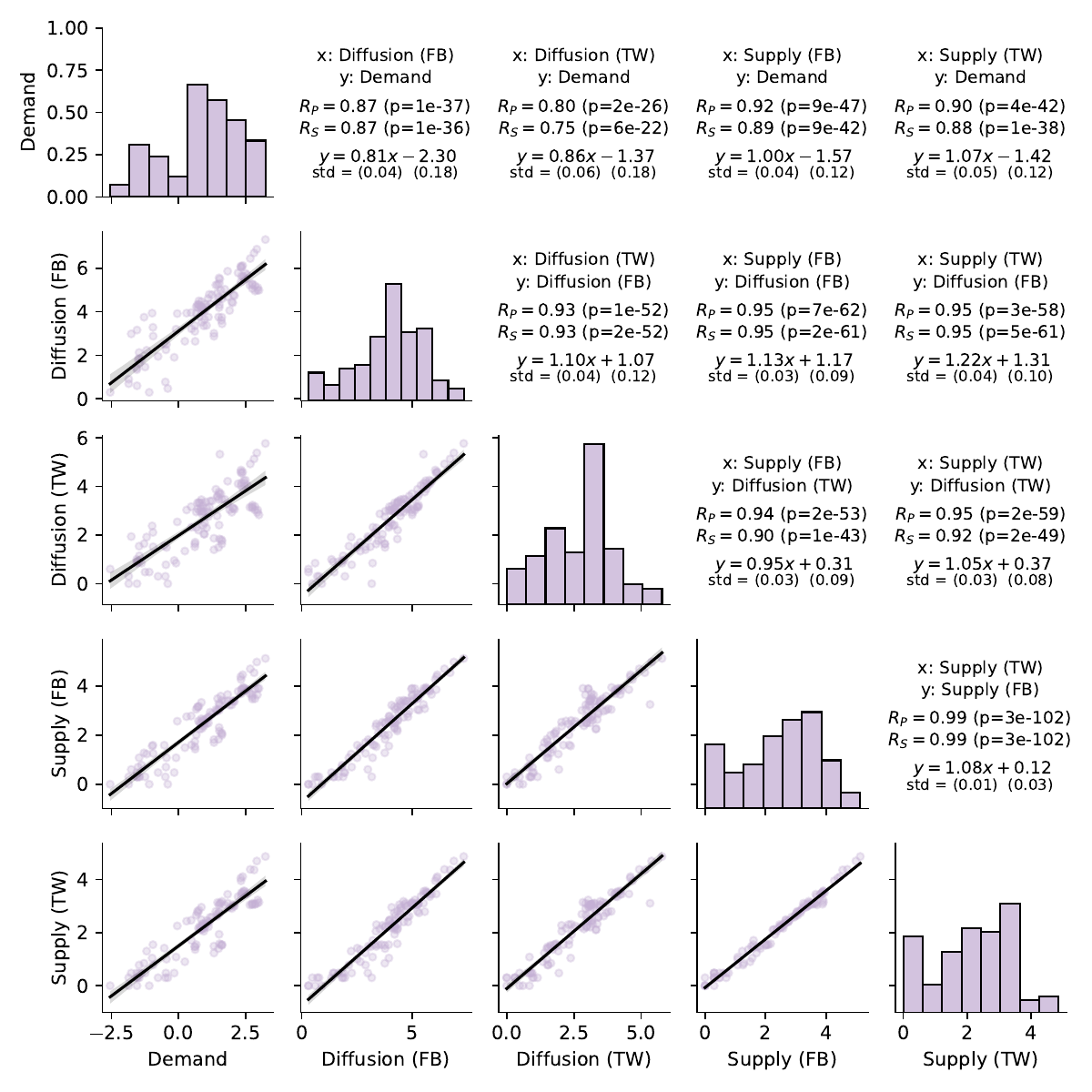}
\caption{Correlation (Pearson and Spearman) and linear regressions between the logarithms of the monthly values of the forces for all different keywords. Supply and Diffusion are reported for both Facebook (FB) and Twitter(TW).}
\label{fig:forces_corr}
\end{figure}

To start studying the relations between the forces, we measure their mutual correlation.
We look at the logarithm of monthly aggregation of the forces in both social networks for all different keywords because.
We report the results in Fig.~\ref{fig:forces_corr}.
In all cases, correlation coefficients are high and significant, and the forces can be considered linearly related.
The monthly aggregation has been chosen to avoid disturbance from the dynamics in shorter windows that might cause larger fluctuations, but similar results are also observed for weekly and daily aggregation, as reported in SI.
Correlation is not causality, so this still does not prove that there are direct relations between the forces.

\begin{figure}[!ht]
\centering
\includegraphics[width=.495\textwidth]{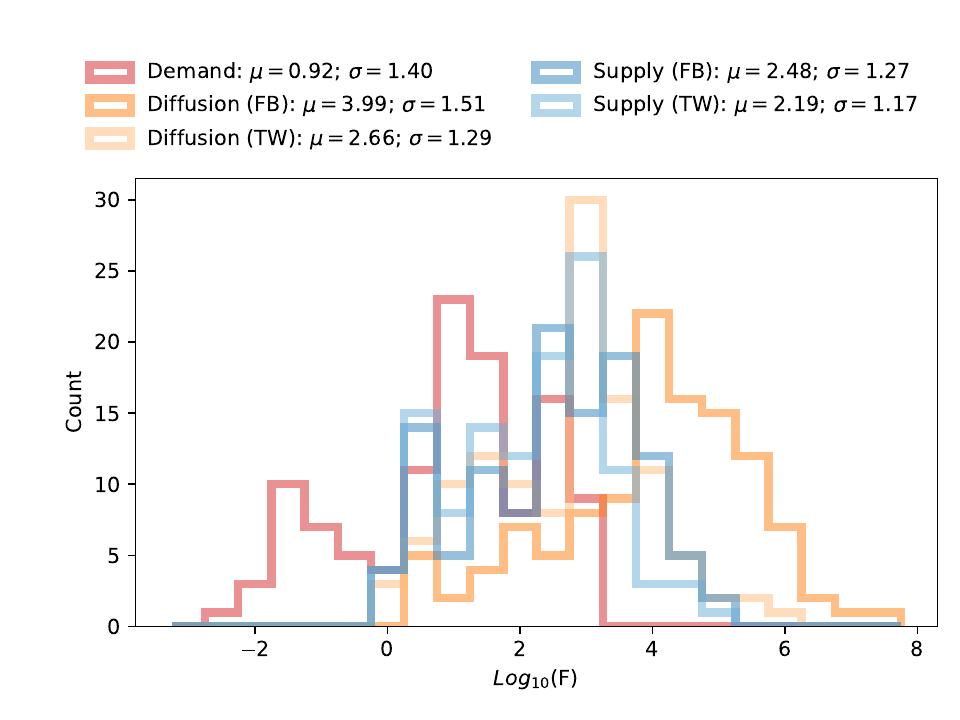}
\includegraphics[width=.495\textwidth]{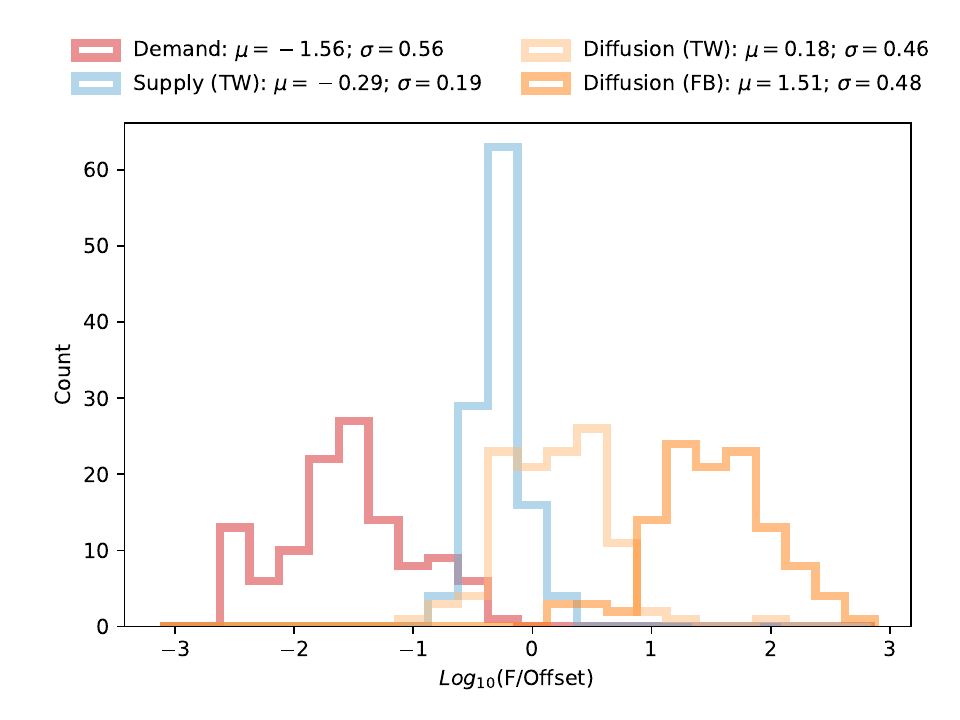}
\caption{\textbf{Left.} The histogram of the monthly values of the forces for all different keywords. \textbf{Right.} The histogram of the monthly values of the forces divided by the Supply (FB) for all different keywords. \textbf{Both.} Supply and Diffusion are reported for both Facebook (FB) and Twitter(TW).}
\label{fig:scale_rel}
\end{figure}

Still, the intercept of the linear regressions of the logarithms suggests that there are typical relative scales that we can measure directly without implying a linear model.
We chose the median force in terms of order of magnitude to be used as an offset (the Supply on Facebook) and normalise by that every monthly aggregate value for each force, for all different keywords.
We report the results in Fig.~\ref{fig:scale_rel}.

For comparison, we also reported the histogram of the monthly aggregated values of the forces. 
We also report the mean and standard values for all histogrammed distributions.
As can be seen by comparing the standard deviations, the scale relations between the forces are much more narrow than the original distributions.
This means that the forces are not only correlated but also refer to dynamics with typical scales tied by precise relation relations.
In other words, if these forces were at equilibrium, given one of them, we could calculate the value that all other forces should have. 
Still, we cannot talk about equilibrium relations if we do not prove dynamic relations between the forces, which is the subject of the next section.

\subsection{Dynamics relations: stationariety and causality}\label{stateq}

\begin{figure}[!ht]
\centering
\includegraphics[width=.495\textwidth]{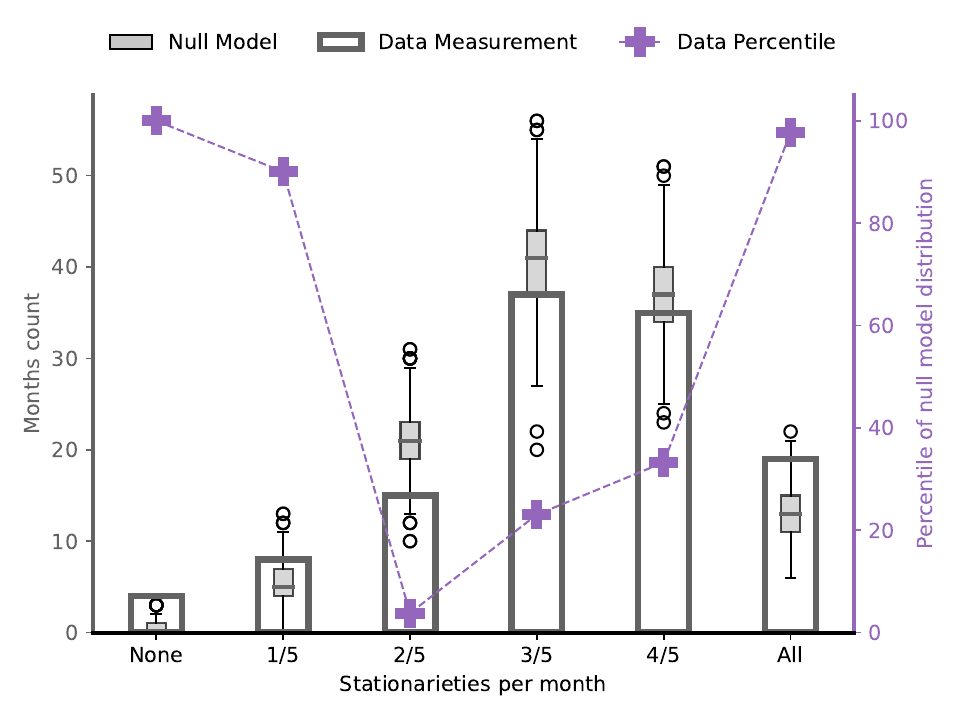}
\caption{The histogram of monthly co-occurring stationarieties of the forces on all months and all keywords, together with a null model obtained by 1000 reshuffling of the monthly sationarieties sequence (the boxplot in grey) and the representation of the actual measured values in terms of percentile (the purple crosses).}
\label{fig:station}
\end{figure}

In order to show that the threee forces trace the same dynamics, we started measuring their stationarity.
The basic idea is that, if the forces are related to the same dynamics, they should be stationary at the same time or non-stationary at the same time. If the forces are not related to the same dynamics, stationarity (or non-stationarity) should not co-occur among different forces more often than in the random case.
We performed the Augmented Dickey-Fuller \cite{Hamilton1994,mackinnon1994approximate,Mushtaq2011,josef_perktold_statsmodelsstatsmodels_2023} on the daily time-series of every force for every keyword and every month.
Then, for every month and every keyword, we measured how many of the five forces were stationary.
The results are reported in Fig.~\ref{fig:station}.
The results must be compared to a null model to understand their significance. 
In fact, if, for example, all forces were stationary for all months and all keywords, stationarieties would obviously always co-occur, but that would not imply any special relation between the forces.
The null model can simply be obtained by reshuffling the sequence of the stationarieties for each force. This will show how many co-occurrences are expected if there is no relation between the forces.
We performed 1000 reshuffling and reported the result in Fig.~\ref{fig:station}, where we also show the percentile of the measurement on the actual data in the distributions obtained from the reshuffling.

\begin{figure}[!ht]
\centering
\includegraphics[width=0.75\textwidth]{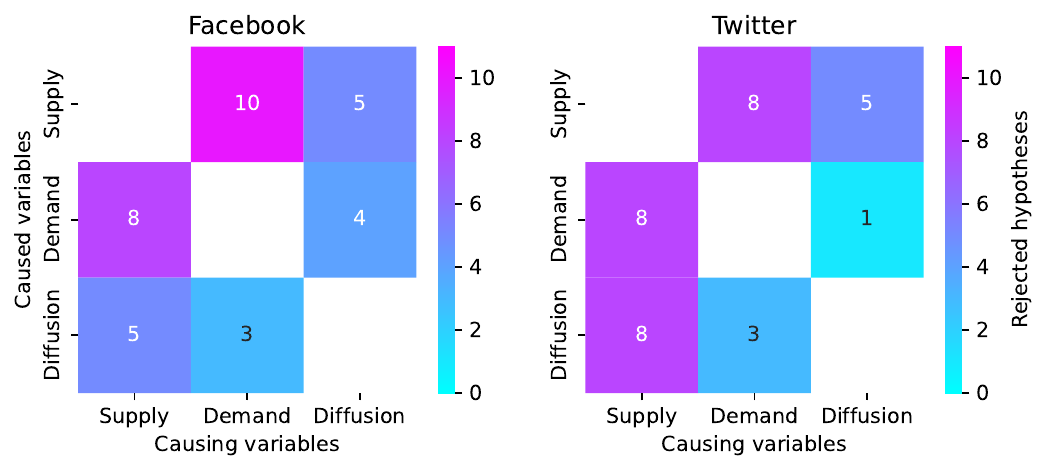}
\includegraphics[width=0.75\textwidth]{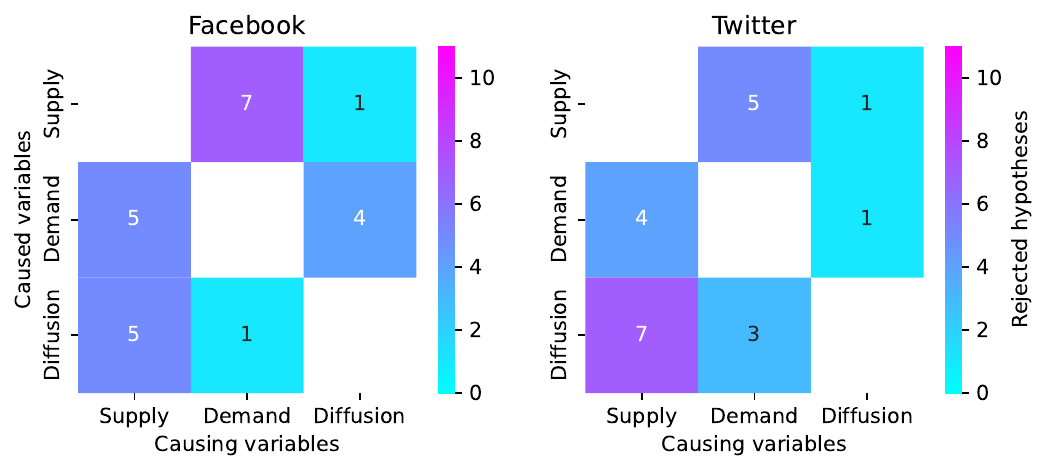}
\caption{Summary of causality tests for information dynamics in Facebook (left panels) and in Twitter (right panels). For the top tables, each cell shows the total number of tests (see Section \ref{matmet}) rejecting the hypothesis that the one variable (cell's column) is causing another (cell's row) conditioned on a third (off cell's axes). For the bottom tables, each cell indicates the total number of rejections from the same tests as above, but counting only the causal effect that is most significant, in the case both directions are rejected (e.g., if conditionally on Diffusion both null hypotheses ``Demand do not causes Supply'' and ``Supply does not cause Demand'' are rejected, only the one with smallest $p$-value contributes to the sum shown in the relative cell).}
\label{fig:causality_tests}
\end{figure}

We observe how the extremes (where none of the five forces or all of them) are much more frequent than in the null model, while the cases in the middle (with around half of the forces stationary) are much less frequent than in the null model.
This suggests the forces are stationary (or non-stationary) all together, so their periods of dynamics seem to be synchronised, as their periods of stasis.
In other words, the forces time-series are telling the story of the same phenomenon from different angles.
To reinforce this conclusion further, we studied the information transfer between the forces to assess their causal relations.
For each keyword, we used a statistical hypothesis test for conditional independence between any two forces time-series conditioning on the third, using a procedure based on resampling via smooth bootstrap.
This allows to assess the significance of information transfer in all six possible direction of causality between the three forces.
For this measure, we considered the two social networks separately, and we report the results in Fig.~\ref{fig:causality_tests}, aggregated by causality direction.

Only a minority of keywords (one for Facebook and four for Twitter) showed no signal.
The most common direction of causality is from Demand to Supply on Facebook, while the importance is more distributed on Twitter. 
Diffusion seems to be the less important in terms of causality, in particular on Twitter.
Still, all combinations and all directions occur.
Fig.~\ref{fig:causality_tests} also reports the comparison between opposite directions, confirming again the importance of the relation between Supply and Demand on Facebook, while on Twitter, the relation from Supply to Demand seems to be more common. 
This relation, present also on Facebook, is expected, to some extent, since Diffusion can happen, by construction, only when Supply is present.
More generally, this analysis shows that the forces exchange information and can drive, in various ways, the system's dynamics.
Now that we have shown that our definitions of the forces are actually tracking the dynamic of the information ecosystem, we speculate that the relative levels shown in Fig.~\ref{fig:scale_rel} are the equilibrium level of the system.
We can now use these forces to assess the status of the health of the system.

\subsection{Semantics and the role of disinformation}\label{disinfo}
\begin{figure}[!ht]
\centering
\includegraphics[width=1\textwidth]{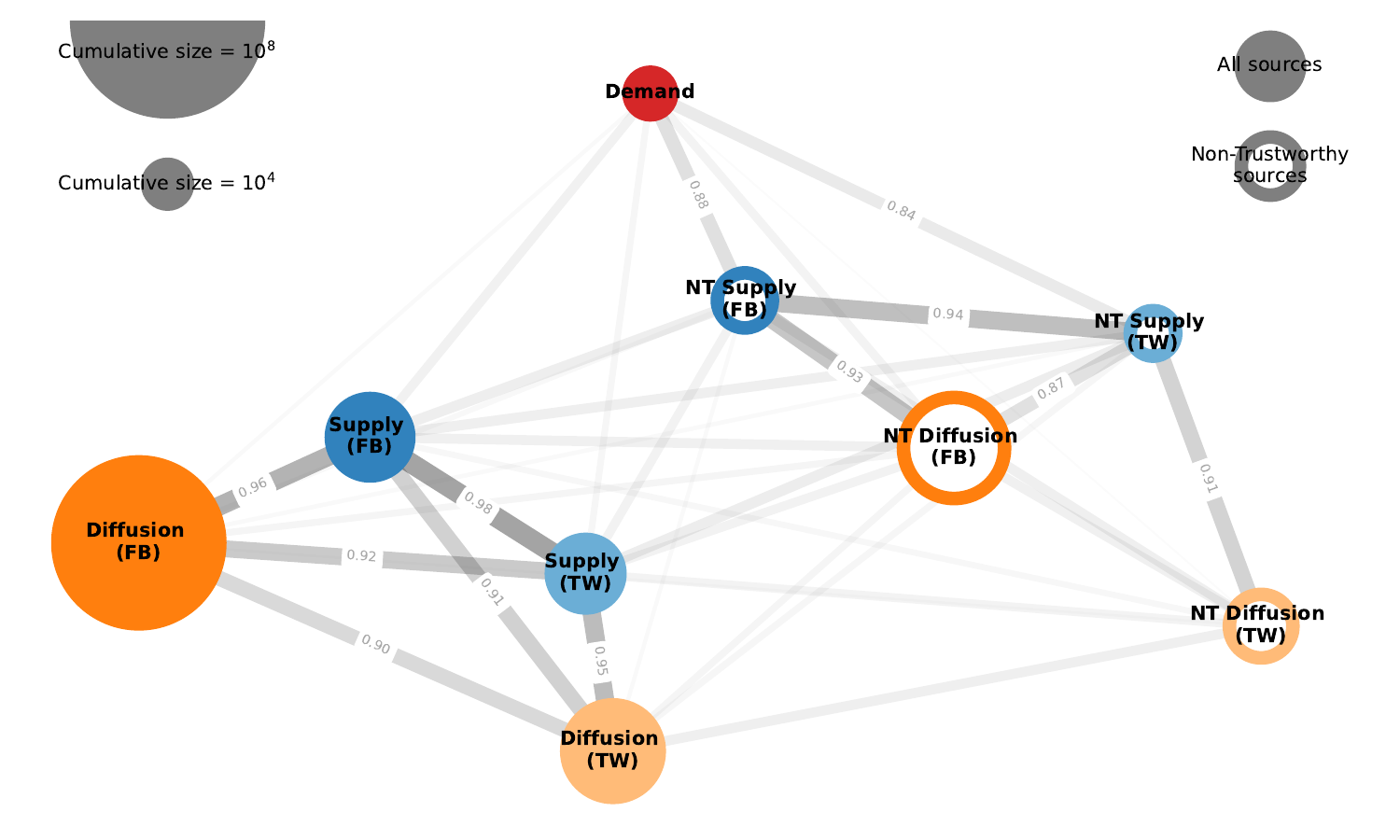}
\caption{The graph of the forces differentiated for the type of sources (all sources and Non-Trustworthy) where the links are the median of the correlations between the daily semantic vectors of the forces. The darker and thicker the link, the higher the value of similarity. Labels are reported for the highest $33\%$ values. The size of the nodes represents the order of magnitude of the cumulative value over the observed period.}
\label{fig:small_graph}
\end{figure}

So far we neglected the semantic aspect, aggregating the different keywords. 
We are now going to define the semantic vectors for each force. 
E.g. the Demand semantic vector for a given day has the values of the demand for the different keywords for that day as components.
The "real" semantic vector would include all possible keywords, while in this work we include only the most important. 
So, even if the conclusions we draw cannot be considered to be valid for the whole system, we can claim that we are studying the most important part of the public debate.
We are also going to analyse disinformation, which in this context we can define for Supply and Diffusion as the production of posts (and their shares) by a subset of sources annotated as "Non-Trustworthy" by professional fact-checkers (see Section \ref{matmet}) on both the analysed social media.
So, now we can define, for each day, nine semantic vectors: one for Demand, four for Supply (for both social media, for all sources and for the Non-Trustworthy subset), and four for Diffusion.
In SI we report a UMAP embedding of all nine vectors for each day. 
We synthesize a more agile representation by calculating the mutual similarities between the vectors for each day.
We chose the Pearsons'correlation similarity between the semantic vectors of the considered couple of forces for a given day.
Then, we took the median of the distances in the daily distribution for each couple of vectors.
All the medians were significant, positive, and high (the minimum observed is $0.66$, between Demand and Diffusion on all sources of Twitter).
Then we arranged the node according to the Fruchterman-Reingold force-directed algorithm \cite{Fruchterman1991,SciPyProceedings_11,noauthor_spring_layout_nodate}, and reported the result in Fig.~\ref{fig:small_graph}.

In this representation, we observe two main features.
First, he two sets of sources (all sources and Non-Trustworthy sources) cluster separately, pointing out that the difference of platform is less important than the type of source.
E.g., Non-Trustworthy Supply on Facebook is closer to Non-Trustworthy Supply on Twitter than on Supply from all sources on Facebook.
This suggests a coherence in the Non-Trustworthy news that transcends the boundaries of the platforms.
Second, perhaps more importantly, the Non-Trustworthy cluster, particularly the Non-Trustworthy supplies, is closer to Demand than the other cluster.
This is a potential danger to social discussion since it suggests that Non-Trustworthy news production and spreading are closer to the community's needs for information than general news.
To conclude the analysis, we also probed the relation between the semantic similarities of general news and the share of Non-Trustworthy news in the overall system.
We measured Spearman's correlation coefficient between the forces similarities in the general news production and the fraction of Non-Trustworthy news Supply and Diffusion over the general news Supply and Diffusion.
We report the results in Table~\ref{tab:ntshare}.
\begin{table}[!ht]
    \centering
    \begin{tabular}{c||c|c|c|c}
        Spearman correlations  & \multicolumn{2}{c|}{ NT Supply Share } &\multicolumn{2}{c}{ NT Diffusion Share } \\
        (* $\Rightarrow p>.05$)&  Facebook & Twitter  & Facebook  & Twitter \\
\hline
\hline
        $R_P(Demand,Supply)$ & $-0.32$ & $-0.57$ & $-0.48$ &  $-0.44$\\
\hline
        $R_P(Diffusion,Supply)$ & $0.01$* &  $-0.31$ & $0.08$* &  $-0.22$\\
\hline
        $R_P(Demand,Diffusion)$ & $-0.41$ &  $-0.57$ & $-0.41$ &  $-0.42$\\
    \end{tabular}
    \caption{The Spearman correlation coefficients of general forces similarities versus the share of Non-Trustworthy supply and diffusion in the two social networks.}
    \label{tab:ntshare}
\end{table}
All the significant coefficients are negative, meaning that the lower the similarities between the forces, the higher the volume of Non-Trustworthy news Supply and Diffusion.
We observe that this effect is limited for the distance between Diffusion and Supply, i.e., it is present only on Twitter.
This is unsurprising since we have already observed how Demand has a central role in the dynamics, especially on Facebook.
More generally, despite the results being an unsettling signal that disinformation providers seem to take advantage of gaps between the forces in general news, they also provide an assessment strategy of the health status of the information market.
In fact, the measurement of the gaps between the forces can show the vulnerability of the information market almost in real-time (daily, at least) to the attempts of escalation in disinformation production and diffusion.
Such a measurement is independent of the direct measure of Non-Trustworthy sources' activity, which usually depends on fact-checkers activity, whose work timescale makes real-time assessment unfeasible.
Our results suggest a possible strategy that can overcome this limitation.

\section{Materials and Methods}\label{matmet}

\subsection{Data collection and pre-elaboration}\label{data}

This study investigates the interplay between the three fundamental forces shaping the news market: Demand, Supply, and Diffusion. 
Namely, we capture Demand through the main terms used in the Google Search Engine from December 2019 to August 2020, as provided by the Google Trends tool. 
The tool does not provide an absolute scale but allows to probe multiple terms simultaneously, returning the results in a scale that can be used for comparison.
In other words, the Demand metric is missing an unknown factor, but this is irrelevant since the comparisons and considerations we draw in the paper do not rely on knowing the absolute scale.
The keywords gathered include:
\begin{itemize}
    \item \textit{Beirut, Campania, Italia, Lombardia, Milano, Piemonte, Roma, Sicilia, Veneto} (terms related to Geographic locations); 
\item \textit{bollettino, casi, contagi, coronavirus, dati, decreto, mappa, morti, sintomi} (terms related to the Covid-19 outbreak);
\item \textit{calcio, campionato, champions, serie A} (terms related to soccer game);
\item \textit{Gioele, Viviana, Viviana Parisi} (terms related to a famous crime news incident in Italy);
\item \textit{Eurovision, Morricone, Sanremo} (terms related to music);
\item \textit{Papa, Papa Francesco} (Pope Francis);
\item \textit{playstation, ps5} (Sony gaming console);
\item \textit{elezioni, regionali, sondaggi} (terms related to political elections);
\item \textit{lotto,  meteo,  news} (other general terms).
\end{itemize}
Many of these terms are semantically overlapping, so we select a shorter set of keywords to account for the most searched topics.
We also remove the Italian locations (except \textit{Rome}) and the term \textit{news} because they are too generic and not related to a specific topic.
The selected list is the following: \textit{beirut}, \textit{calcio}, \textit{coronavirus}, \textit{elezioni}, \textit{Eurovision}, \textit{lotto}, \textit{meteo}, \textit{Morricone}, \textit{Papa}, \textit{playstation}, \textit{Roma}, \textit{Sanremo}, \textit{sondaggi}, \textit{Viviana}.
To analyse the news Supply, we rely on a list of news outlets provided by AGCOM, the Italian Authority for Communications Guarantees, which covers the main leaders of information in Italy during the time span under analysis~\cite{agcom2018}. 
The list includes traditional newspapers, online-only news outlets, information agencies, TV, radio websites, and scientific sources. 
Moreover, the data have specific annotations on Non-Trustworthy sources.
The source-based methodology is widely recognized and firmly established in the existing literature on disinformation~\cite{grinberg2019}. 
We adopt the same method, which is especially suitable for examining the conduct of Non-Trustworthy sources, as in the current study.
Note that the resulting leader dataset is the same used in~\cite{gravino2022} to which the reader can refer for further details.

Limited to contents containing the selected keywords, we use news posted by the selected news outlets on the two major social media platforms in Italy — Facebook and Twitter — as a proxy for the Supply. 
We trace the Diffusion of these contents through the corresponding user engagement, represented by the number of shares a post gained on the belonging platform.
For gathering data from Facebook, we rely on CrowdTangle~\cite{crowdtangle2023}, a Facebook-owned tool that tracks interactions on public content from various social media platforms. For Twitter, we exploit the official API accessed through the academic account before the limitations introduced by the new management\footnote{https://twitter.com/XDevelopers/status/1621026986784337922}.

The final supply dataset consists of $2,112,678$ Facebook posts and $1,410,711$ tweets from $411$ different news sources, as more clearly detailed in Table~\ref{tab:data_breakdown}, which in turn also includes the corresponding Diffusion statistics.

\begin{table}[ht]
    \centering
    \begin{tabular}{c||c|rr|rr}
         & Sources & \multicolumn{2}{c|}{Facebook} & \multicolumn{2}{c}{Twitter} \\
         & & posts & shares & tweets & retweets\\
         \hline
         \hline
         Reliable & $330$ & $1,956,941$ & $121,108,643$ & $1,302,008$ & $6,291,579$\\
         Non-Trustworthy & $81$ & $155,737$ & $11,225,099$ & $108,703$ & $2,102,646$
    \end{tabular}
    \caption{Breakdown of the Supply and Diffusion dataset. Data are divided by source type (Reliable or Non-Trustworthy) and social media platform (Facebook or Twitter).}
    \label{tab:data_breakdown}
\end{table}

In addition to being functional for a swift identification of Diffusion statistics, using social media production as a proxy for Supply by selected information leaders is also legitimate, given the high and significant correlation it has with the general direct production of news from the same news sources~\cite{gravino2022}, as reported in Table~\ref{tab:data_corr}.
\begin{table}[ht]
    \centering
    \begin{tabular}{c||cc|cc}
          & \multicolumn{2}{c|}{Facebook} & \multicolumn{2}{c}{Twitter} \\
         & Corr & $p$-value & Corr & $p$-value\\
         \hline
         \hline
         Pearson $R_p$ & $0.791$ & $\sim 0$ & $0.861$ & $\sim 0$\\
         Spearman $R_s$ & $0.783$ & $\sim 0$ & $0.872$ & $\sim 0$
    \end{tabular}
    \caption{Correlation (Pearson and Spearman) between the daily direct supply of news from the selected information leaders and the corresponding social media production on Facebook and Twitter, respectively.}
    \label{tab:data_corr}
\end{table}

In our work, we performed different kinds of aggregation on different series.
The Demand is already provided by the Google Trends platform as a daily time-series.
Supply and Diffusion have also been aggregated daily to have a similar format.
Then, for the different analyses, different levels of aggregation have been adopted: daily, monthly or over the whole period. 
The level of aggregation used for every analysis is described in the main text.

\subsection{Causal analysis}\label{causal_analysis}
\begin{figure}[!ht]
\centering
\includegraphics[width=0.65\textwidth]{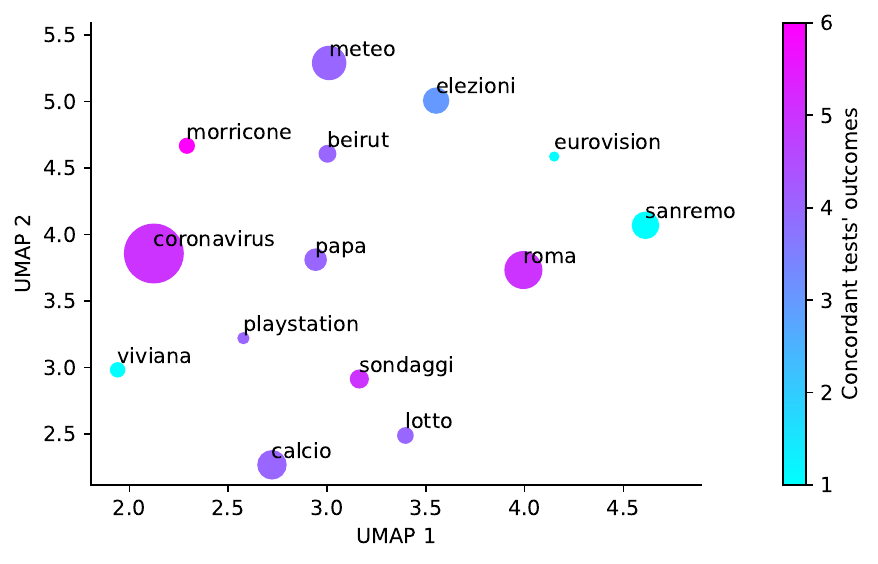}
\caption{Embedding of keywords' test significance. For every keywords, the vector of the $p$-values from the twelve tests for causal analysis, from Facebook and Twitter dynamics, is projects in two dimensions via UMAP embedding using the cosine distance. Each dot is coloured by the number of tests (out of six) that are rejected or accepted for both social media. Dots are sizes according the square root of their total supply.}
\label{fig:test_embedding}
\end{figure}

To investigate causal interactions between time-series of Supply, Demand and Diffusion of a given keyword in a given social media (Fig.~\ref{fig:causality_tests} and ~\ref{fig:test_embedding}), we used a statistical hypothesis test for conditional independence between time-series based on resampling via smooth bootstrap~\cite{efron1994introduction,prevedello2024cittest}.
This testing procedure is a nonparametric counterpart of the Granger causality test that relaxes the requirements imposed by vector autoregression modelling and improves upon other nonparametric techniques based on local permutation resampling~\cite{runge2018conditional}.

Briefly, to test for the null hypothesis that two signals $X$ and $Y$ are independent conditioned on the signal $Z$, the Transfer Entropy statistic is used~\cite{schreiber2000measuring}. To approximate its distribution, the statistic is calculated over data resampled via smooth bootstrap:
first, the joint distribution of $X_t,\ldots,X_{t-m}$, $Y_t,\ldots,Y_{t-m}$, $Z_t,\ldots,Z_{t-m}$ is estimated by Kernel Density Estimation (KDE) with Gaussian kernels from the observations $D = (x_t,\ldots,x_{t-m}$, $y_t,\ldots,y_{t-m}$, $z_t,\ldots,z_{t-m})_{t \geq m}$, with lag $m=7$ days, using Scott's bandwidth, and imposing that the covariance between $X$ and $Y$ given $Z$ is null;
from this distribution, the dataset $D$ is sampled $B=1000$ times, each time drawing from the KDE as many samples as observations, thus generating $D^*_1, \ldots, D^*_B$;
finally, the statistic $S$ is calculated on every bootstrapped dataset to determine the $p$-value $p = B^{-1} \sum_{i=1}^B \mathbf{1}(\{ S(D^*_i)>=S(D) \})$, where $\mathbf{1}(A)=1$ if $A$ is true and is null otherwise.

For each keyword and social media, six $p$-values were then obtained, one for every permutation of (Supply, Demand, Diffusion) assigned to $(X, Y, Z)$. Each set of six $p$-values was adjusted for multiple testing comparisons using the Holm-Bonferroni method.

\section{Conclusions}\label{conc}

The digital age provides new challenges as information travels more quickly in a system of increasing complexity. 
But it also provides new opportunities, as we can more easily track and study digital trails of the system.
These trails have often been studied separately focusing on different aspects (like disinformation production or sharing dynamics) individually.
In this work, we propose to study the news ecosystem as an information market by analysing three main metrics: Supply, Demand, and Diffusion of information.
Working on a dataset relative to Italy from December 2019 to August 2020, we validate the choice of the metrics, proving their static and dynamic relations.
We demonstrate that they seem to have specific equilibrium relative levels.
We reveal the strategic role of Demand in leading a non-trivial network of causal relations.
We show how disinformation news Supply and Diffusion seem to cluster by transcending social media platforms.
It also appears to be closer to information Demand than the general news Supply and Diffusion, implying a potential danger to the health of the public debate.
Finally, we prove that the share of disinformation in the Supply and Diffusion of news has a significant linear relation with the gap between Demand and Supply/Diffusion of news from all sources.

This work confirms and expands the result of a previous work \cite{gravino2022}, pointing out the potential of the analysis of the whole news ecosystem through its main drivers.
The results proved to be valid for different keywords and on different social media platforms.
This is another step toward a potential real-time analysis of the information market to assess and possibly prevent vulnerabilities.
Still, the work presents limitations, and there is much more to do.
The analyses need to be expanded to different countries and different timeframes.
Also, a more comprehensive strategy for keyword selection could be defined, e.g. leveraging topic detection algorithms.
On the theoretical side, the challenge is on one side to define and test a model to describe the complex behaviour of the system, replicating all the crucial aspects.
Such a model should be able to reproduce, for example, what happens when the equilibrium relative levels are modified by external perturbations.
On the other side, besides the disinformation fraction, a wider set of metrics (e.g. polarisation) can be taken into account to assess the status of the system's health.
The comparative analysis of the model metrics and the semantic relations could then be crossed with these metrics to be able to perform a more detailed real-time evaluation of the news ecosystem vulnerabilities.
\backmatter

\bmhead{Supplementary information}

\begin{figure}[!ht]
\centering
\includegraphics[width=.95\textwidth]{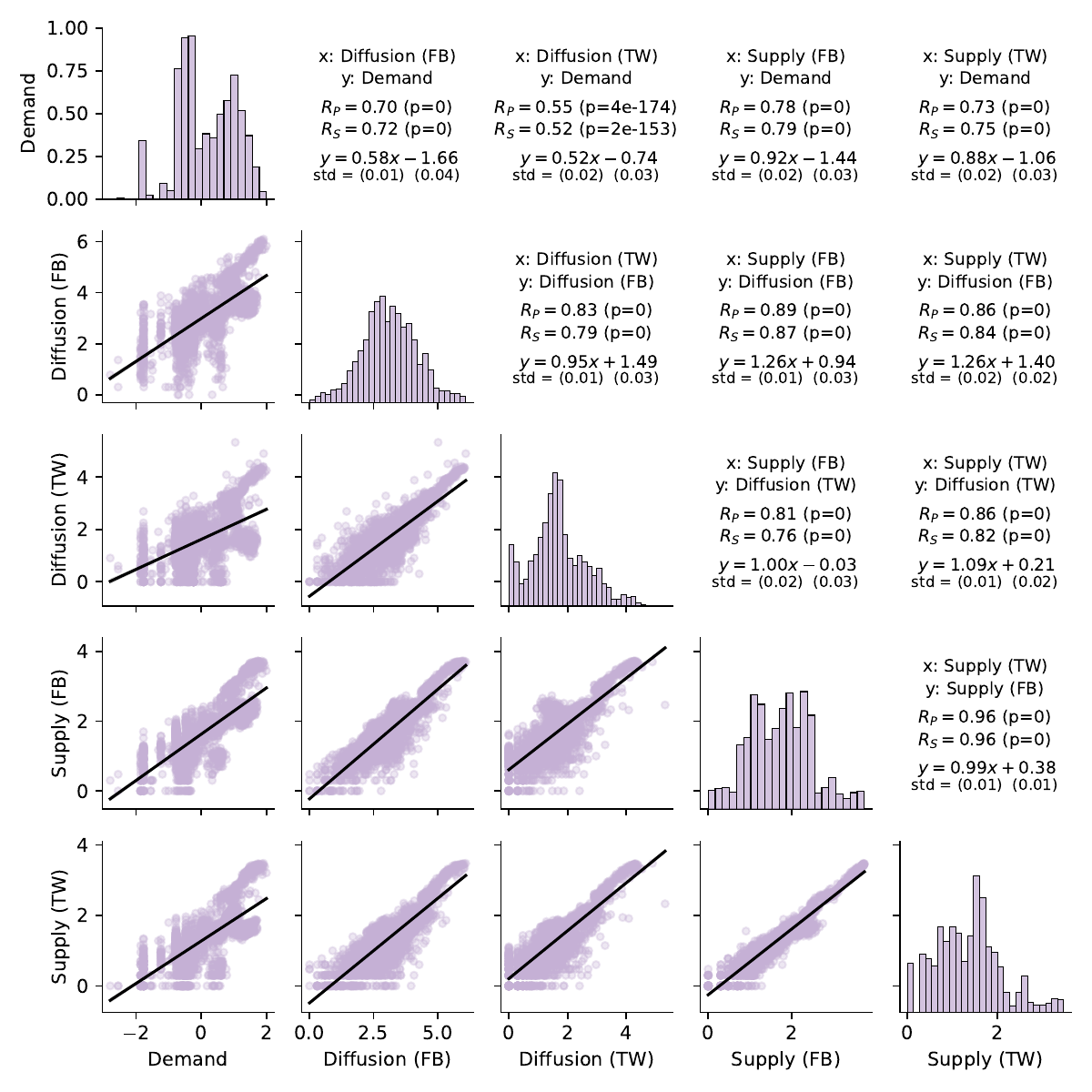}
\caption{Correlation (Pearson and Spearman) and linear regressions between the logarithms of the daily values of the forces for all different keywords. Supply and Diffusion are reported for both Facebook (FB) and Twitter(TW).}
\label{fig:forces_corr_daily}
\end{figure}

\begin{figure}[!ht]
\centering
\includegraphics
[width=.95\textwidth]{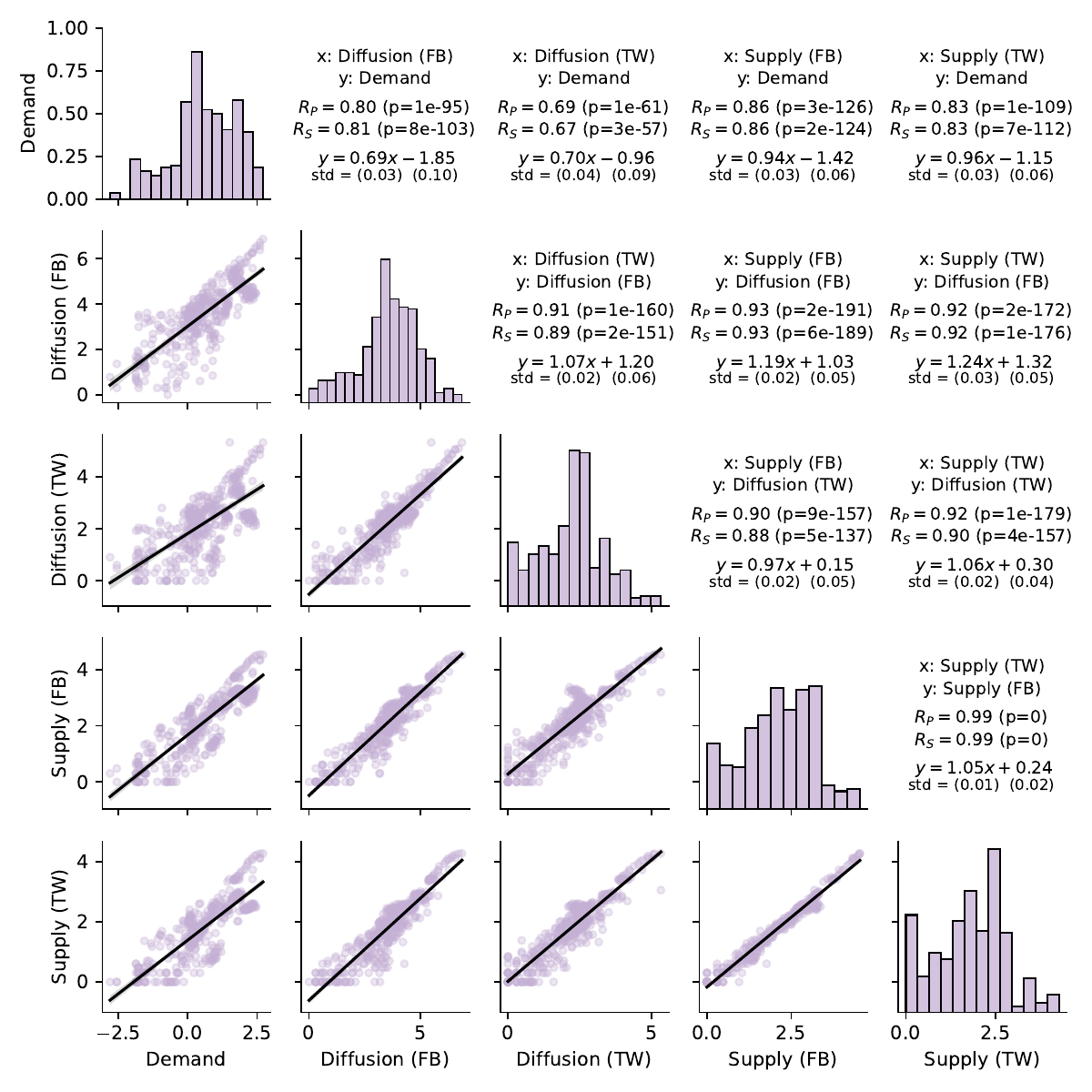}
\caption{Correlation (Pearson and Spearman) and linear regressions between the logarithms of the weekly values of the forces for all different keywords. Supply and Diffusion are reported for both Facebook (FB) and Twitter(TW).}
\label{fig:forces_corr_weekly}
\end{figure}

\begin{figure}[!ht]
\centering
\includegraphics[width=1.\textwidth]{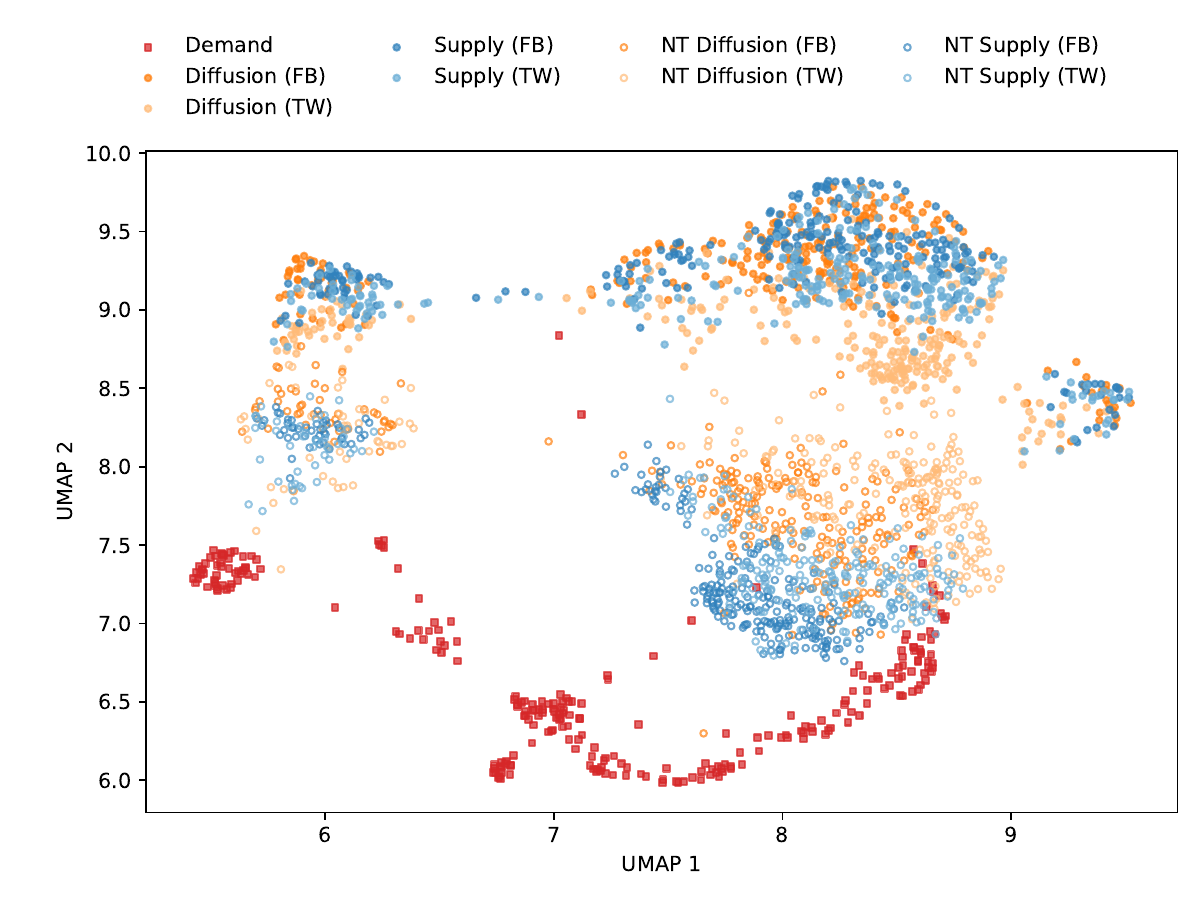}
\caption{2-D UMPA embedding using the correlation distance of the forces daily semantic vectors, differentiated for the type of sources: all sources and Non-Trustworthy (NT).}
\label{fig:umap_days}
\end{figure}

\newpage
\newpage

\bmhead{Acknowledgments}

We thank M. Delmastro of AGCOM for providing access to the database of Italian news outlets. 
The database was shared in the framework of the Task Force on ‘Digital Platforms and Big Data - Covid-19 Emergency’, established by AGCOM to contribute, among other things, to the fight against online disinformation on issues related to the COVID-19 crisis. 


\bmhead{Declarations}


\begin{itemize}
\item Funding

This work has been supported by the Horizon Europe VALAWAI project (grant agreement number 101070930).

\item Competing interests

The authors declare no competing interests.

\item Ethics approval 

Not applicable

\item Consent to participate

Not applicable

\item Consent for publication

Not applicable

\item Availability of data and materials

Google Search engine data were generated by the Google Trends platform and is publicly available at \url{https://trends.google.com}. 
Derived data for Supply, Demand and Diffusion supporting the findings of this study are available at \url{https://github.com/SonyCSLParis/news_searches}.

\item Code availability 

All codes for data analysis are available at \url{https://github.com/SonyCSLParis/news_searches}.
\item Authors' contributions

Conceptualization: Gravino. Methodology: Gravino, Prevedello. Validation: Gravino, Prevedello. Software: Gravino, Prevedello. Writing-original draft: Gravino, Prevedello, Brugnoli. Visualization: Gravino, Prevedello. Supervision: Gravino. Proofread: Gravino, Prevedello, Brugnoli. All authors read and approved the final manuscript.

\end{itemize}


\bigskip









\bibliography{sn-bibliography}

\end{document}